# Metadata salad at the Cordoba Observatory

Veronica Lencinas[1]

[1]*Observatorio Astronómico. Universidad Nacional de Córdoba*

**Abstract**

The Plate Archive of the Cordoba Observatory includes 20.000 photographs and spectra on glass plates dating from 1893 to 1983. This contribution describes the work performed since the plate archive was transferred to the Observatory Library in 2011. In 2014 an interdisciplinary team was assembled and a research grant from the National University of Cordoba was obtained with the objectives of preserving the glass plates and generate public access for astronomers and other audiences. The preservation work not only includes practical intervention to improve conservation conditions for the whole archive, but also a diagnose of the preservation conditions for the plates and identification of best practices for cleaning the plates. The access envisioned through digitization requires not only the scanning of all the plates, but also careful definition and provision of metadata. In this regard, each institutional level involved – in this case: archive, library, astronomical observatory and public university - demands and provides different bibliographic practices, involving multiple standards of description and coding.

## Introduction

The Córdoba Observatory was founded in 1871 as Argentine National Observatory (Observatorio Nacional Argentino). Domingo F. Sarmiento and the U.S. astronomer Benjamin A. Gould met in 1865 in Boston where Gould expressed to Sarmiento his intentions to create a complete catalog of the southern sky. Once elected president, Sarmiento offered Gould the creation and direction of a new scientific institution: the Argentine National Observatory in the mediterranean city of Córdoba[1]. Benjamin A. Gould would not reach completely his ambitious goal: his "Uranometría Argentina" was indeed the first complete star catalog of the Southern hemisphere, but the observations had to be made without instruments since the Franco-Prussian War blocked the shipping of all equipment. After the war, Gould and his staff continued their work and published several works, for example the Zone Catalog (1884) and the General Argentine Catalog (1886). Gould returned in 1885 to the United States. But Benjamin A. Gould was also a pioneer of astronomy photography. His "Cordoba Photographs: photographic observations of star-clusters", published in 1897, were the first systematic use of photography in astronomy. The original plates of the "Cordoba Photographs" on wet collodion are actually part of the Harvard Plate Collection.

In 1954 the Argentine National Observatory was transferred to the National University of Córdoba and changed its name to Astronomical Observatory of Córdoba (Observatorio Astronómico de Córdoba).

## The Plate Archive of the Cordoba Observatory

The Plate Archive of the Observatory includes ca. 20.000 photographs and spectra on dry plates created between 1893 and 1983. From 1880 on, dry plates with silver bromide emulsions replaced other photographic mediums like albumen and collodion plates. The main advantages of dry plates were the feasibility of industrial manufacture and their stability who allowed storage and distribution while maintaining photographic permanence [2]. The Plate Archive of the Córdoba Observatory is organized in series responding to their scientific origin like Astrographic Catalog, Carte du Ciel, Pulkovo Zones, Sun, Planets, Comets and Minor Bodies, Globular Clusters, Magellanic Clouds, Southern Galaxies and others. In addition there are also photographs of the Observatory, its construction, staff, instruments, activities and the Fonds of spectra [3]. The sizes of the plates varies from 5.5 x 8.5 cm to approximately 20 x 30 cm. Most of them have data written on their borders and have been stored in their original boxes. Only the most recent plates from the 40's on have cardboard enclosures with some data about the plate. The Observatory kept also observation log books with additional information. The location of the Plate Archive in a location below ground without ventilation and summer temperatures up to 32º C / 89.6º F and 80% moisture was far from optimal. Since neither the enclosures nor the boxes met the requirements for long term conservation, through the project "Recovery, value enhancement and dissemination of the heritage of astronomical photographs of the Plate Archive of the Astronomical Observatory of Córdoba" the replacement of all enclosures and boxes has started. The main goals of the project are the preservation of the plates and to generate access for astronomers and other public through digitization and metadata creation. In 2016 the Plate Archive was relocated to a climate controlled storage.

## Creating access (Metadata salad)

The Plate Archive was organized in 1996 and most plates recorded in a basic inventory. A group of astronomers under the leadership of Jesus Calderon nearly completed a catalog of astronomical plates. In 2012 the Archive was transferred to the Observatory library. Digitization of 500 plates started in 2012 through a grant from the Bunge and Born Foundation. In 2015 the Observatory acquired a scanner to digitize the whole Archive. Regardless of the digitized images, to publish these images to end users, first the the recipients of the images have to be defined and with it their requirements regarding descriptive metadata.

The Plate Archive is first of all an archive, a "... documentary by-product of human activity retained for their long-term value"[4]. As an archive the application of standards of archival organization and document description follows established and approved practices. On the second instance, the Córdoba Plate Archive is a special collection of a library and that means that its records have to coexist with norms, practices and software used by the library without losing its identity as archive. It is also part of an astronomical observatory and its natural users are astronomers who in turn have their own standards of identification and data description. But astronomers are not the only recipients of the photographs; as a part of a national University -a public institution- the Plate Archive needs also to be included in the outreach and scientific dissemination activities of the Observatory and offer final results to the general public. Each one of the institutional levels: archive, library, observatory and university, have their own ideas, expectations and standards on how information and data should be organized and displayed. It is also important to keep in mind that the original photograph (plate) and the digital image are two different documents, one derived from the other and that both need a proper and specific description even if they share part of the data.

The term "metadata" is formed by a combination of the greek μετα, meta, that means means "after" or "beyond", and the latin "datum", "something given" and is used in this combination with the meaning "data about data." According to Taylor & Joudrey, the concept of metadata refer to "structured information that describes the attributes of information resources for the purposes of identification, discovery, selection, use, access, and management"[5].

### 1. Metadata in archives

Archivists have their own theoretical-practical corpus that is different from other documentary disciplines like librarianship or museology. One of their basic principles is the "Respect des fonds" who unites two basic principles of archival organization. The principle of provenance requires that all records created, assembled or maintained by an organization or an individual are represented together and distinguishable from the records of any other organization or individual. The principle of original order means that the order of the records that was established by the creator should be maintained by physical and/or intellectual means whenever possible to preserve existing relationships between the documents and the evidential value inherent in their order. Both principles form the basis for archival descriptions [6]. These principles requires to group documents and record their data according the order in they are found or reconstruct the original order and represent the data consistent with these principles. The physical organization of an archive is structured from the most general to the most specific, identifying and creating hierarchical relations between different levels of arrangement. Metadata for archives need to recognize and represent these hierarchies.

One of the pillars of archival description is the International Standard Archival Description (General), in short ISAD(G), published by the International Council of Archives in 2000. ISAD(G) is a standard widely used in Argentina since there are no national archival standards. ISAD(G) includes five levels of arrangement: fonds, subfonds, series, subseries, file and item [7].

For an archive that is part of a library, there is another standard of interest: "Describing Archives: a Content Standard" (DACS) is the official standard of the Society of American Archivists. DACS is the U.S. embodiment of ISAD(G) and is also compatible with the Anglo American Cataloging Rules and Resource Description Access (RDA), both library standards. DACS offers the documentalist much more detail and data fields to construct an adequate description than ISAD(G). DACS offers also "crosswalks" between DACS, RDA and Marc21, a library coding standard. In spite of the benefits of DACS for archives who are part of a library, there is no Spanish translation available who could promote its use in Spanish speaking countries.

DACS recognizes and recommends other more standards for specific materials. For archives with graphic materials, including photographs it advocates "Descriptive Cataloging of Rare Materials (Graphics)" (DCRM(G)). This norm was published in 2013 by the Rare Book and Manuscript Section de la Association of College and Research Libraries. For a Plate Archive, chapter 5 "Physical Description Area", is of special interest because it offers rules to create more detailed descriptions of the medium and their material. A wealth of examples are also included [8].

## 2. Metadata in libraries

Actually most libraries in Argentina still use the Anglo American Cataloging Rules, second edition, while in other countries the transition to the new cataloging code, "Resource Description and Access" (RDA) is underway. Even though the Anglo American Cataloging Rules include a chapter about manuscripts that is applicable to archives, it is RDA who has incorporated much more elements of archival practice for example access point for families and devised titles.

Libraries have also a large tradition of coding rules. The Marc 21 Bibliographic Format (former Usmarc Format) is an important international standard supported by most common bibliographic utilities. In the United States, archives started to incorporate their records in library public access catalogs in the 80's in spite of the more limited nature of library catalog data in comparison to archival finding aids. But their inclusion in library catalogs facilitated the access of archival documents for end users. The most important problem for archives who use library coding formats is the poor aptitude of Marc21 to handle hierarchical relationships and to group records. But in practice the limitations to represent multilevel relations are mostly originated in library software packages more than the coding format itself.

## 3. Metadata in observatories

Astronomy is a scientific discipline with its own tradition of publication and sharing of primary data like images and spectra. The universal standard for digital images in astronomy is the FITS (Flexible Image Transport System) format. Unlike other metadata, the FITS format includes embedded metadata and is structurally very different to file formats used in the archival and library community like Marc21 or XML based formats.

As a very specialized file format, FITS is not useful for outreach or dissemination purposes. For these activities, more generally used file formats like JPEG or PNG and conventional content management systems or document repository software are much more suited. As an institution belonging to a national University who was a pioneer in including outreach as one of its fundamental goals, the Observatory has an important outreach and dissemination activity. The library participates in these activities and propose to publicize the images of the Plate Archive for a general public through the institutional repository of the National University of Córdoba. This requires careful planning of file format conversion, define resolution and size of the images to be displayed and also the creation of appropriate metadata for a more diverse user community. The repository of the University uses the DSpace software and in spite of its ability to accommodate different metadata schemes, the use of the Dublin Core Scheme will facilitate the dissemination of the information through the OAI-PMH protocol.

## Metadata processing for several kinds of users

Faced with a variety of applicable norms and standards, for the Plate Archive of the Córdoba Observatory three lines of action were defined. 1) Describe the photographic plates with the compatible standards DACS and RDA and complement the physical description with DCRM(G). This decision allow the implementation of Marc21, the use of standard library software and the creation of links between plate data and publications. 2) Use of the FITS format for the digital images of the Archive and offer access to the astronomical community through the Nuevo Observatorio Virtual Argentino (NOVA), a member of the International Virtual Observatory Alliance (IVOA). 3) Batch loading of selected images of some plates for the general public in the institutional repository of the University in a standard file formats (JPEG or PNG) and with a document description in the Dublin Core Schema.

All metadata in the Archive starts with a base record who includes text written on the plate, information recorded on the enclosure, box or log books and the observation of the documentalist. This base record is then converted to the secondary formats: Marc21 for the library system and the public catalog, FITS header and Dublin Core record for the repository. At this point a automatized procedure exists for the Marc21 format.

Is is important to reclaim Plate Archives as an important scientific but also cultural heritage. Using standards permits incorporation, sharing and reuse of digital data through different networks: astronomical, libraries, archives and other heritage organizations for a broad and diverse user community.

Special thanks to Dr. Diego García Lambas (Director of the Observatory), Dr. Martín Leiva (Outreach Secretary), GAF (Astrometry and Photometry Group), NOVA (Argentine Virtual Observatory), Santiago Paolantonio and Edgardo Minniti (authors and consultants), volunteers, interns, researchers and many others.

## References

[1] Minniti, E. R., y Paolantonio, S. (2009). Córdoba Estelar. Historia Del Observatorio Nacional Argentino. Universidad Nacional de Córdoba.
[2] Lavédrine, B. (2009). Photographs of the past: process and preservation. Los Angeles, Calif: Getty Conservation Institute.
[3] Calderón, J. H., Fierro, I. B., Melia, R., Willimoës, C., & Giuppone, C. (2004). The Digital Archive of the Photographic Images of the Córdoba Observatory Plates Collections. Astrophysics and Space Science, 290(3-4), 345–351.
[4] International Council of Archives. All you have ever wanted to know about archives and ICA… http://www.ica.org/en/all-you-have-ever-wanted-know-about-archives-and-ica
[5] Taylor, A. G., & Joudrey, D. N. (2009). The organization of information (3rd ed.). Westport: Libraries Unlimited.
[6] Society of American Archivists. (2013). Describing archives: a content standard (2nd ed.). Chicago: Society of American Archivists.
[7] International Council on Archives. (2000). ISAD(G): General International Standard Archival Description. Ottawa: ICA.
[8] Rare Books and Manuscripts Section of the Association of College and Research Libraries (2013). Descriptive Cataloging of Rare Materials (Graphics). Chicago: ACRL.